\documentclass{ifacconf}

\usepackage{graphicx}      
\usepackage{natbib}        

	\usepackage{amsmath,amsfonts,amssymb,mathbbol}
	\usepackage{enumerate}
	\usepackage{psfrag}         
	\usepackage{pstool} 				

\newcommand{\ys}{y}

\newcommand{\ppp}{\mathit{p}}
\newcommand{\qqq}{\mathit{q}}
\newcommand{\rrr}{\mathit{r}}

\newcommand{\dif}{\mathrm{d}}

\newcommand{\inv}{\ensuremath{^{-1}}} 

\newcommand{\Iden}[1]{\ensuremath{\mathcal{I}_{#1}}} 

\newcommand{\real}{\mathcal{R}} 
\newcommand{\reals}[1]{\real^{#1}}

\begin{document}
\begin{frontmatter}

\title{PI(D) tuning for Flight Control Systems via Incremental Nonlinear Dynamic Inversion}

\author[dlr]{Paul~Acquatella B.\thanksref{footnotedlr}}
\author[dlr,tud]{Wim van Ekeren\thanksref{footnoteitud1}}
\author[tud]{Qi Ping Chu\thanksref{footnoteitud2}}

\address[dlr]{DLR, German Aerospace Center\\
Institute of System Dynamics and Control\\
D-82234 Oberpfaffenhofen, Germany}
\address[tud]{Delft University of Technology, Faculty of Aerospace Engineering \\ 2629HS Delft, The Netherlands}

\thanks[footnotedlr]{Research Engineer, Space Systems Dynamics Department. \texttt{paul.acquatella@dlr.de}.}
\thanks[footnoteitud1]{Graduate Student, Control \& Operations Department. \texttt{w.vanekeren@student.tudelft.nl}.}
\thanks[footnoteitud2]{Associate Professor, Control \& Operations Department. \texttt{q.p.chu@tudelft.nl}.}

	\begin{abstract}
	Previous results reported in the robotics literature show the relationship
	between \textit{time-delay control} (TDC) and \textit{proportional-integral-derivative control} (PID).
	In this paper, we show that incremental nonlinear dynamic inversion (INDI) --- more familiar in the aerospace community ---
	are in fact equivalent to TDC. This leads to a meaningful and systematic method for PI(D)-control tuning of 
	robust nonlinear flight control systems	via INDI. We considered a reformulation of the plant dynamics inversion
	which removes effector blending models from the resulting control law, 
	resulting in robust model-free control laws like PI(D)-control. 
	\end{abstract}

\end{frontmatter}

\section{Introduction}\label{intro}
Ensuring stability and performance in between operational points of 
widely-used gain-scheduled linear PID controllers motivates the use of nonlinear dynamic inversion (NDI) for flight control systems.
NDI cancels out nonlinearities in the model via state feedback, and then linear control
can be subsequently designed to close the systems' outer-loop, hence eliminating the need
of linearizing and designing different controllers for several operational points as in gain-scheduling. 

In this paper we consider nonlinear flight control strategies based on incremental nonlinear dynamic inversion (INDI). 
Using sensor and actuator measurements for feedback allows
the design of an incremental control action which, in combination with nonlinear dynamic
inversion, stabilizes the \textit{partly}-linearized nonlinear system \textit{incrementally}. 
With this result, dependency on exact knowledge of the system dynamics is greatly reduced, 
overcoming this major robustness issue from conventional nonlinear dynamic inversion. 
INDI has been considered a sensor-based approach because sensor measurements 
were meant to replace a large part of the vehicle model.

Theoretical development of increments of nonlinear control action date back
from the late nineties and started with activities concerning `implicit dynamic inversion'
for inversion-based flight control (\cite{Smith1, Bacon1}), where the architectures considered in
this paper were firstly described. Other designations for these developments found
in the literature are `modified NDI' and `simplified NDI', but the designation `incremental
NDI', introduced in (\cite{Chen}), is considered to describe the methodology and nature of these type
of control laws better (\cite{Chen, Chu2, Sieberling2010}). INDI has been elaborated and applied theoretically
in the past decade for advanced flight control and space applications (\cite{Sieberling2010, Smith1, Bacon1, Bacon2, Bacon3, Acquatella2012, Simplicio2013}). 
More recently, this technique has been applied also in practice for quadrotors and adaptive control (\cite{Smeur2016b, Smeur2016a}).

In this paper, we present three main contributions in the context of nonlinear flight control system design.

1) We revisit the NDI/INDI control laws and we establish the equivalence between INDI and \textsl{time-delay control} (TDC).

2) Based on previous results reported in the robotics literature showing the relationship
between discrete formulations of TDC and \textsl{proportional-integral-derivative control} (PID), 
we show that an equivalent PI(D) controller with gains $<K$, $T_i$, $(T_d)>$  tuned via 
INDI/TDC is more meaningful and systematic than heuristic methods,
since one considers \textsl{desired} error dynamics given by Hurwitz gains $<k_P$, $(k_D)>$. Subsequently, 
tuning the remaining effector blending gain is much less cumbersome than designing a whole set of gains iteratively.

3) We also consider a reformulation of the plant dynamics inversion as it is done in TDC
which removes the effector blending model (control derivatives) from the resulting control law. 
This has not been the case so far in the reported INDI controllers, causing robustness problems
because of their uncertainties. Moreover, this allows to consider 
the introduced term as a scheduling variable which is only directly related to the proportional gain $K$.

\section{Flight Vehicle Modeling}\label{ndi}
We are interested in Euler's equation of motion representing flight vehicles' angular velocity dynamics
\begin{equation}\label{euler1}
	I\,\dot{\omega}+\omega\times I\,\omega = M_B
\end{equation}
where $M_B\in\reals{3}$ is the external moment vector in body axes,
$\omega\in\reals{3}$ is the angular velocity vector,
and $I\in\reals{3\times 3}$ the inertia matrix of the rigid 
body assuming symmetry about the plane $x-z~$ of the body.

Furthermore, we will be interested in the time history of the angular velocity vector,
hence the dynamics of the rotational motion of a vehicle \eqref{euler1}
can be rewritten as the following set of differential equations
\begin{equation}\label{dynamics1}
	\dot{\omega} = I^{-1}\big(~M_B-\omega\times I\,\omega~\big)
\end{equation}
where
\begin{align*}
	\omega =
	\left[ 
		\begin{array}{c} 
					\ppp \\ 
					\qqq \\ 
					\rrr \\
		\end{array}
	\right], &\quad
	M_B =
	\left[ 
		\begin{array}{c} 
					\mathit{L} \\ 
					\mathit{M} \\ 
					\mathit{N} \\
		\end{array}
	\right] =
	S Q
	\left[ 
		\begin{array}{c} 
					b\mathit{C_l} \\ 
					\overline{c}\mathit{C_m} \\ 
					b\mathit{C_n} \\
		\end{array}
	\right],	\\
	I &= 
	\left[
		\begin{array}{ccc} 
					~I_{xx}~~  & 0~~       & I_{xz}~ \\ 
					~0~~       & I_{yy}~~  & 0~ \\ 
					~I_{xz}~~  & 0~~       & I_{zz}~
		\end{array} 
	\right],
\end{align*}
with $\ppp, \qqq, \rrr, $ the body roll, pitch, and yaw rates, respectively; 
$\mathit{L}, \mathit{M}, \mathit{N}, $ the roll, pitch, and yaw moments, respectively; 
$S$ the wing surface area, $Q$ the dynamic pressure, $b$ the wing span, $\overline{c}$ the mean
aerodynamic chord, 
and $\mathit{C_l}, \mathit{C_m}, \mathit{C_n}$ the moment coefficients for roll, pitch, and yaw, respectively.
Furthermore, let $M_B$ be the sum of moments partially generated by the
aerodynamics of the airframe $M_a$ and moments generated by control surface 
deflections $M_c$, and we describe $M_B$ linearly in the 
deflection angles $\delta$ assuming the control derivatives
to be linear as in~\cite{Sieberling2010} with $(M_c)_\delta = \frac{\partial}{\partial\delta}M_c$; therefore
\begin{equation}
		M_B = M_a + M_c = M_a + (M_c)_\delta\delta
\end{equation}
where
\begin{equation*}
	M_a =
	\left[ 
		\begin{array}{c} 
					\mathit{L_a} \\ 
					\mathit{M_a} \\ 
					\mathit{N_a} \\
		\end{array}
	\right],~
	M_c =
	\left[ 
		\begin{array}{c} 
					\mathit{L_c} \\ 
					\mathit{M_c} \\ 
					\mathit{N_c} \\
		\end{array}
	\right],~		
	\delta =
	\left[ 
		\begin{array}{c} 
					\delta_a \\ 
					\delta_e \\ 
					\delta_r \\
		\end{array}
	\right] 
\end{equation*}	
and $\delta$ corresponding to the control inputs: aileron, elevator, and rudder deflection angles, respectively.
Hence the dynamics \eqref{dynamics1} can be rewritten as
\begin{equation}\label{dynamics2}
	\dot{\omega} = f(\omega) + G(\omega)\delta
\end{equation}
with
\begin{align*}
	f(\omega) = I^{-1}\big(M_a - \omega\times I\,\omega\big),~
	G(\omega) = I^{-1}(M_c)_\delta.
\end{align*}

For practical implementations, we consider first-order actuator dynamics represented by the following transfer function
\begin{equation}\label{eq:actdyn}
	\frac{\delta}{\delta_c} = G_a(s) = \frac{K_a}{\tau_a s + 1},
\end{equation}
and furthermore, we do not consider these actuator dynamics in the control design process as it is usually
the case for dynamic inversion-based control. For that reason, we assume that these actuators 
are \textit{sufficiently fast} in the control-bandwidth sense, meaning that $1/\tau_a$ is higher than the
control system closed-loop bandwidth.

\section{Flight Control Law Design}
\label{sec:fcld}

\begin{figure*}[!t]
	\centering   	
	\vspace{-20mm}
	\psfrag{988}{\tiny Reference}
	\psfrag{987}{\tiny Trajectory}  
	\psfrag{990}{\tiny Position}
	\psfrag{989}{\tiny Control}
	\psfrag{993}{\tiny ~Flight Path}
	\psfrag{992}{\tiny ~Angle and}
	\psfrag{888}{\tiny ~Airspeed Control}        
	\psfrag{996}{\tiny Attitude}
	\psfrag{995}{\tiny Control}
	\psfrag{998}{\tiny Rate Control}
	\psfrag{991}{\tiny $X,Y,Z$}
	\psfrag{994}{\tiny $V, \psi, \gamma$}
	\psfrag{997}{\tiny $\mu, \alpha, \beta$}
	\psfrag{999}{\tiny $p,q,r$}            
	\includegraphics[width=0.9\textwidth]{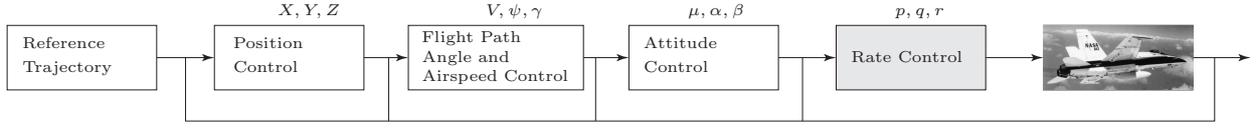}
	\caption
	{Four loop nonlinear flight control design. 
	We are focused on nonlinear dynamic inversion of the rate control loop (grey box) in the following. Image credits:~\cite{Sonn}.}
	\label{fig:loops}
\end{figure*}	
	
\subsection{Nonlinear Dynamic Inversion}

Let us define the control parameter to be the angular velocities, 
hence the output is simply $y = \omega$.
We then consider an error vector defined as $e = y_d - y$
where $y_d$ denotes the \textit{smooth} desired output vector (at least one time differentiable).
	
Nonlinear dynamic inversion (NDI) is designed to linearize and decouple the 
rotational dynamics in order to obtain an \textit{explicit} desired closed loop dynamics
to be followed. Introducing the virtual control input $\nu = \dot{\omega}_{\text{des}}$,
if the matrix $G(\omega)$ is non-singular (i.e., invertible) 
in the domain of interest for all $\omega$, 
the nonlinear dynamic inversion control consists in 
the following input transformation (\cite{Slotine, Chu2})
\begin{equation}\label{eq:input-transformation-MIMO}
	\delta = G(\omega)\inv\big[\nu - f(\omega)\big]	
\end{equation}
which cancels all the nonlinearities, 
and a simple input-output linear relationship 
between the output $y$ and the new input $\nu$ is obtained as
\begin{align}\label{eq:linear-MIMO}
	\dot{y} = \nu 
\end{align}
Apart from being linear, an interesting result from this relationship 
is that it is also decoupled since the input $\nu_i$ only affects the 
output $\ys_i$. 
From this fact, the input transformation \eqref{eq:input-transformation-MIMO} is called a \textit{decoupling control law},
and the resulting linear system \eqref{eq:linear-MIMO} is called the \textit{single-integrator} form.
This single-integrator form \eqref{eq:linear-MIMO} can be rendered exponentially stable with
\begin{align}\label{eq:virtual-control-law-error}
	\nu = \dot{y}_{d} + k_P e
\end{align}
where $\dot{y}_{d}$ is the feedforward term for tracking tasks, and
$k_P\in\reals{3\times 3}$ a constant diagonal matrix, whose $i-$th
diagonal elements $k_{P_i}$ 
are chosen so that the polynomials
\begin{align}\label{eq:error-poly}
	s + k_{P_i} \qquad (i = p, q, r)
\end{align}
may become Hurwitz, i.e., $k_{P_i}<0$. This results in the exponentially stable and decoupled \textit{desired} error dynamics
\begin{align}\label{eq:error-dyn}
	\dot{e} + k_P e = 0
\end{align}
which implies that $e(t)\rightarrow 0$.	
From this typical tracking problem it can be seen that the entire 
control system will have two control loops (\cite{Chu2, Sieberling2010}): 
the inner linearization loop \eqref{eq:input-transformation-MIMO},
and the outer control loop \eqref{eq:virtual-control-law-error}.	 
This resulting NDI control law depends on accurate knowledge
of the aerodynamic moments, hence it is 
susceptible to model uncertainties contained in both $M_a$ and $M_c$.

In NDI control design, we consider outputs with relative degrees of one (rates), meaning a first-order system to be controlled, see Fig.~\ref{fig:loops}.
Extensions of input-output linearization for systems involving higher relative degrees are done via \textit{feedback linearization}
(\cite{Slotine, Chu2}).
			
\subsection{Incremental Nonlinear Dynamic Inversion}\label{indi}

The concept of incremental nonlinear dynamic inversion (INDI) amounts to the application of NDI 
to a system expressed in an incremental form. This improves the robustness of the closed-loop 
system as compared with conventional NDI since dependency on the accurate 
knowledge of the plant dynamics is reduced. Unlike NDI, this control design technique 
is \textit{implicit} in the sense that desired closed-loop dynamics do not reside in 
some explicit model to be followed but result when the feedback loops are closed (\cite{Bacon1, Bacon2}).

To obtain an incremental form of system dynamics, we
consider a first-order Taylor series expansion of $\dot{\omega}$ 
(\cite{Smith1, Bacon1, Bacon2, Bacon3, Sieberling2010, Acquatella2012, Acquatella2013}),
not in the geometric sense, but with respect to a \textit{suffiently small} time-delay $\lambda$ as  
\begin{align*}\label{eq:taylor-incremental}
	\dot{\omega}  
		= ~ & \dot{\omega}_0 + 
		\frac{\partial}{\partial\omega}\big[f(\omega) + G(\omega)\delta\big] \bigg|_{\substack{\omega=\omega_0 \\ \delta=\delta_0}}\left(\omega-\omega_0\right) \\
		& + \frac{\partial}{\partial\delta}\big[G(\omega)\delta\big] \bigg|_{\substack{\omega=\omega_0 \\ \delta=\delta_0}}\left(\delta-\delta_0\right) 
			+ \mathcal{O}(\Delta\omega^2, \Delta\delta^2) \\ 
		\cong ~ & \dot{\omega}_0 + f_0\left(\omega-\omega_0\right) + G_0\left(\delta-\delta_0\right)
\end{align*}
with
\begin{subequations}\label{eq:MIMO-dynamics-incremental-derivatives}
	\begin{align}
		\dot{\omega}_0 &\equiv f(\omega_0)+G(\omega_0)\delta_0 = \dot{\omega}(t-\lambda)
	\end{align}
\end{subequations}
where $\omega_0 = \omega(t-\lambda)$ and $\delta_0 = \delta(t-\lambda)$ are
the time-delayed signals of the current state $\omega$ and control $\delta$, respectively.
This means an approximate linearization about the $\lambda-$delayed signals is performed \textit{incrementally}. 

For such sufficiently small time-delay $\lambda$ so 
that $f(\omega)$ does not vary significantly during $\lambda$,
we assume the following approximation to hold
\begin{equation}\label{eq:eps-indi}
	{\epsilon_{INDI}}(t) \equiv f(\omega(t-\lambda)) - f(\omega(t)) \cong 0
\end{equation}
which leads to
\begin{equation}\label{eq:inc-odot}
	\Delta\dot{\omega} \cong G_0\cdot\Delta\delta
\end{equation}
Here,
$	\Delta\dot{\omega} = \dot{\omega}-\dot{\omega}_0 = \dot{\omega}-\dot{\omega}(t-\lambda)$
represents the incremental acceleration, and
$	\Delta\delta = \delta-\delta_0$
represents the so-called incremental control input.
For the obtained approximation $\dot{\omega} \cong \dot{\omega}_0+G_0(\delta-\delta_0)$, 
NDI is applied to obtain a relation between 
the incremental control input and the output of the system
\begin{equation}\label{eq:incremental-second-relation}
	\delta =\delta_0 + G_0^{-1}\big[\nu-\dot{\omega}_0\big]	
\end{equation}

Note that the deflection angle $\delta_0$ that corresponds to $\dot{\omega}_0$ is 
taken from the output of the actuators, and it has been assumed that a 
commanded control is achieved \textit{sufficiently fast} according to the assumptions of
the actuator dynamics in \eqref{eq:actdyn}. The total control command along with the obtained 
linearizing control $\Delta\delta$ can be rewritten as
\begin{equation}\label{eq:incremental-fifth-relation}
	\delta(t) = \delta(t-\lambda) + G_0^{-1}\Big[\nu-\dot{\omega}(t-\lambda)\Big].
\end{equation}

The dependency of the closed-loop system on accurate knowledge of the airframe model in $f(\omega)$
is largely decreased, improving robustness against model uncertainties contained therein.
Therefore, this implicit control law design is more dependent 
on accurate measurements or accurate estimates of $\dot{\omega}_0$, the angular acceleration,
and $\delta_0$, the deflection angles, respectively. 

{\textsl{\textbf{Remark 1}}}:
By using the measured $\dot{\omega}(t-\lambda)$ and $\delta(t-\lambda)$ incrementally
we practically obtain a robust, model-free controller with the self-scheduling properties of NDI.

Notice, however, that typical INDI control laws are nevertheless also depending 
on effector blending models reflected in $G_0$, which makes this implicit controller 
susceptible to uncertainties in these terms.
Instead, consider the following transformation as in (\cite{Chang2009})
\begin{equation}\label{dyngbar}
	\dot{\omega} = H + \bar{g}\cdot\delta 	
\end{equation}
with
\begin{equation*}
	H(t) =  f(\omega) + (G(\omega) - \bar{g})\delta,
\end{equation*}
and with the following (but not limited) options for $\bar{g}$ (\cite{Chang2009}),
where $n=3$ in our case
\begin{equation*} 
	\bar{g}_1 = k_G\cdot\Iden{n} = 
	k_G
	\left[
		\begin{array}{cccc} 
			1 & 0 & \cdots & 0 \\
			0 & 1 &  &  \\
			\vdots & & \ddots \\ 
			0 & & &  1
		\end{array}
	\right],
	~
	\bar{g}_2 =
	\left[
		\begin{array}{cccc} 
			k_{G_1} & 0 & \cdots & 0 \\
			0 & k_{G_2} & &  \\
			\vdots & &  \ddots\\ 
			0 & & & k_{G_n}
		\end{array}
	\right].
\end{equation*}		

Applying nonlinear dynamic inversion (NDI) to \eqref{dyngbar} results
in an expression for the control input of the vehicle as
\begin{equation}\label{eq:ndi-new}
	\delta(t) = \bar{g}\inv\big[\nu(t) - H(t)\big].
\end{equation}

Considering $H_0 = \dot{\omega}_0 - \bar{g}\cdot\delta_0$, 
the incremental counterpart of \eqref{eq:ndi-new}
results in a control law that is neither depending on the airframe model nor the effector blending moments 
\begin{equation}\label{eq:indi-new}
	\delta(t) = \delta(t-\lambda) + \bar{g}^{-1}\Big[\nu-\dot{\omega}(t-\lambda)\Big].	
\end{equation}

{\textsl{\textbf{Remark 2}}}:
The self-scheduling properties of INDI in \eqref{eq:incremental-fifth-relation} due
to the term $G_0$ are now lost, suggesting that $\bar{g}$ should be an scheduling variable.
	
\subsection{Time Delay Control and Proportional Integral control}

\textit{Time delay control} (TDC) (\cite{Chang2009}) departs from the usual dynamic inversion input transformation
of \eqref{dyngbar}
\begin{equation} 
	\delta(t) = \bar{g}\inv\big[\nu(t) - \bar{H}(t)\big]
\end{equation}		
where $\bar{H}$ denotes an estimation of $H$, being the nominal case when $\bar{H}=H$ which results
in perfect inversion. Instead of having an estimate, the TDC takes the following assumption (\cite{Chang2009})
analogous to \eqref{eq:eps-indi}
\begin{equation} 
	{\epsilon_{TDC}}(t) \equiv H(t-\lambda)-H(t) \cong 0.
\end{equation}		

This relationship is used together with \eqref{dyngbar} to obtain what is called
\textit{time-delay estimation} (TDE) as the following
\begin{equation}
	\bar{H} = H(t-\lambda) = \dot{\omega}(t-\lambda) - \bar{g}\cdot\delta(t-\lambda)	
\end{equation}		

In addition, ${\epsilon}(t)$ is called TDE \textit{error} at time $t$. Combining 
the equations we obtain the following TDC law
\begin{equation}\label{eq:indi-new-vis}
	\delta(t) = \delta(t-\lambda) + \bar{g}^{-1}\big[\nu-\dot{\omega}(t-\lambda)\big]
\end{equation}		
which is in fact \textit{equivalent} to the INDI control law obtained in \eqref{eq:indi-new}.
Appropriate selection of $\bar{g}$ must ensure stability according to (\cite{Chang2009}),
and ideally, this term should be tuned according to the best estimate of the true effector 
blending moment $\hat{g}(\tilde{\omega})$ for measured angular velocities $\tilde{\omega}$.	

So far we have considered derivations in continuous-time. 
For practical implementations of these controllers and for the matters of upcoming discussions,
sampled-time formulations involving continuous and discrete quantities as in (\cite{Chang2009}) are more convenient
and restated here. 
For that, considering that the smallest $\lambda$ one can consider is the equivalent of the 
sampling period $t_s$ of the on-board computer. The sampled formulation of \eqref{eq:indi-new-vis} may be expressed as
\begin{equation} 
	\delta(k) = \delta(k-1) + \bar{g}^{-1}\big[\nu(k-1)-\dot{\omega}(k-1)\big]
\end{equation}		
where it has been necessary to consider $\nu$ at sample $k-1$ for causality reasons.
Replacing the sampled virtual control $\nu$ according to \eqref{eq:virtual-control-law-error} we have
\begin{equation}\label{eq:disc-indi}
		\delta(k) = \delta(k-1) + \bar{g}^{-1}\big[\dot{e}(k-1) + k_P e(k-1)\big]			
\end{equation} 
and we can consider the following finite difference approximation of the error derivatives
as angular accelerations are not directly measured
\begin{align}
	\dot{e}(k) = [e(k)-e(k-1)]/t_s.
\end{align}

Consider now the standard \textit{proportional-integral} (PI) control
\begin{equation} 
	\delta(t) = K\big(e(t)+T_I^{-1}\int_0^t e(\sigma)\dif\sigma\big) + \delta_{DC},	
\end{equation}		
where $K\in\reals{3\times 3}$ denotes a diagonal proportional gain matrix,
$T_I\in\reals{3\times 3}$ a constant diagonal matrix representing a reset or integral time,
and $\delta_{DC}\in\reals{3}$ denotes a constant vector representing a trim-bias,
which acts as a trim setting and is computed by evaluating the initial conditions. 
The discrete form of the PI is given by
\begin{equation} 
	\delta(k) = K\big(e(k-1)+T_I^{-1}\sum_{i=0}^{k-1}t_s e(i)\big) + \delta_{DC}	
\end{equation}		

When substracting two consecutive terms of this discrete formulation, we can remove the 
integral sum and achieve the so-called PI controller in incremental form
\begin{equation}\label{eq:disc-pi}
	\delta(k) = \delta(k-1) + K\cdot t_s\big(\dot{e}(k-1) + T_I^{-1}\cdot e(k-1)\big)		
\end{equation}

Following the same steps, and for completeness, we also present the PID extension by simply considering the
extra derivative term $\ddot{e}$ 
\begin{equation*}\label{eq:disc-pid}
	\delta(k) = \delta(k-1) + K\cdot t_s\big(T_D\ddot{e}(k-1) + \dot{e}(k-1) + T_I^{-1}\cdot e(k-1)\big),		
\end{equation*}
where $T_D\in\reals{3\times 3}$ denotes a constant diagonal matrix representing derivative time.

\subsection{Equivalence of INDI/TDC/PI(D)}

Having in mind the found the equivalence between INDI and TDC, and
comparing terms from \eqref{eq:disc-indi} with \eqref{eq:disc-pi}, 
we have the following relationships as originally found in (\cite{Chang2009}) which are 
the relationship between the discrete formulations of TDC and PI in incremental form
\begin{align}
	K = (\bar{g}\cdot t_s)^{-1}, \quad T_I = k_P^{-1}
\end{align}

Whenever the system under consideration is of second-order controller canonical form, we
will have error dynamics of the form $\ddot{e} + k_D\dot{e} + k_P e = 0$, and considering
the newly introduced derivative gain $k_D$ related to $\ddot{e}$ we have
\begin{align}
	K =  k_D\cdot (\bar{g}\cdot t_s)^{-1}, \quad T_I =  k_D\cdot k_P^{-1}, \quad T_D =  k_D^{-1}
\end{align}

This suggests not only that an equivalent discrete PI(D) controller with gains $<K$, $T_i$, $(T_d)>$ 
can be obtained via INDI/TDC, but doing so is more meaningful and systematic than heuristic methods.
This is because we begin the design from \textsl{desired} error dynamics given by Hurwitz gains $<k_P$, $(k_D)>$ 
and what follows is finding the remaining effector blending gain $\bar{g}$ either analytically whenever $G$ is well known,
with a proper estimate $\hat{G}$, or by tuning according to closed-loop requirements. 
As already mentioned, details on a sufficient condition for closed-loop 
stability under discrete TDC, and therefore applicable to its equivalent INDI, can be found in (\cite{Chang2009}) and the references therein.

In essence, this procedure is more efficient and much less cumbersome than designing a whole set of gains iteratively. Moreover,
for flight control systems, the self-scheduling properties of inversion-based controllers have suggested superior advantages
with respect to PID controls since these must be gain-scheduled according to the flight envelope variations. The relationships here outlined
suggests that PID-scheduling shall be done at the proportional gain $K$ via the effector blending gain $\bar{g}$, and \textit{not} over 
the whole set of gains $<K$, $T_i$, $(T_d)>$.
%
\section{Longitudinal flight control simulation}
\label{sec:launcher}

	In this section, robust PI tuning via INDI is demonstrated
	with a simple yet significant example consisting of the tracking control design for a longitudinal
	launcher vehicle model. The second-order nonlinear model is obtained from (\cite{Sonn,Kim2004}), and 
	it consists on longitudinal dynamic equations representative of a vehicle
	traveling at an altitude of approximately 6000 meters, with aerodynamic coefficients
	represented as third order polynomials in angle of attack $\alpha$ and Mach number $M$.
	
	The nonlinear equations of motion in the pitch plane are given by
	\begin{subequations}\label{eq:example}
		\begin{align}
			\dot{\alpha} &= q + \frac{\bar{q}S}{m V_T}\bigg[C_z(\alpha,M)+b_z(M)\delta\bigg],\\
			\dot{q} &= \frac{\bar{q}S d}{I_{yy}}\bigg[C_m(\alpha,M)+b_m(M)\delta\bigg],
		\end{align}
	\end{subequations}	%
	where
	\begin{align*}
		C_z(\alpha,M) &= \varphi_{z1}(\alpha)+\varphi_{z2}(\alpha)M, \\
		C_m(\alpha,M) &= \varphi_{m1}(\alpha)+\varphi_{m2}(\alpha)M, \\
		b_z(M) &= 1.6238M-6.7240, \\
		b_m(M) &= 12.0393M-48.2246,
	\end{align*}
	and
	\begin{align*}
		\varphi_{z1}(\alpha) &= -288.7\alpha^3+50.32\alpha\left|\alpha\right|-23.89\alpha, \\
	  \varphi_{z2}(\alpha) &= -13.53\alpha\left|\alpha\right|+4.185\alpha, \\
		\varphi_{m1}(\alpha) &= 303.1\alpha^3-246.3\alpha\left|\alpha\right|-37.56\alpha, \\
	  \varphi_{m2}(\alpha) &= 71.51\alpha\left|\alpha\right|+10.01\alpha.
	\end{align*}		

	These approximations are valid for the flight envelope of $-10^\circ\leq\alpha\leq10^\circ$
	and $1.8\leq M \leq 2.6$. To facilitate the control design, the nonlinear longitudinal
	model is rewritten in the more general state-space form as
	\begin{subequations}
		\begin{align}
			\dot{x}_1 &= x_2 + f_1(x_1) + g_1 u \label{forceeq} \\ 
			\dot{x}_2 &= f_2(x_1) + g_2 u
		\end{align}
	\end{subequations}	
	where:
	\begin{align*}
		x_1 &= \alpha, &
		\quad
		x_2 &= q \\
		g_1 &= C_1 b_z, &
		\quad
		g_2 &= C_2 b_m
	\end{align*}
	and
	\begin{align*}
		f_1(x_1) &= C_1\big[\varphi_{z1}(x_1)+\varphi_{z2}(x_1)M\big],& \quad C_1 &= \frac{\bar{q}S}{m V_T}, \\
		f_2(x_1) &= C_2\big[\varphi_{m1}(x_1)+\varphi_{m2}(x_1)M\big],& \quad C_2 &= \frac{\bar{q}S d}{I_{yy}}.
	\end{align*}

	The control objective considered here is to design a PI autopilot
	via INDI that tracks a smooth command reference $y_r$ with the pitch rate
	$x_2$. It is assummed that the aerodynamic force and moment functions
	are accurately known and the Mach number $M$ is treated as a parameter
	available for measurement. 
	Moreover, for this second-order system in non-lower triangular form due to $g_1 u$ and $f_2(x_1)$, 
	pitch rate control using INDI is possible due to the time-scale separation principle (\cite{Chu2, Sieberling2010}).
	With respect to actuator dynamics modeled as in \eqref{eq:actdyn}, we consider $K_a = 1$, and $\tau_a = 1e^{-2}$.
	
	\subsection{Pitch rate control design}
		
	First, introduce the rate-tracking error
		\begin{align}
			z_2 &= x_2-x_{2_{ref}} 
		\end{align}
	the $z_2-$dynamics satisfy the following error
		\begin{align}
			\dot{z}_2 &= \dot{x}_2-\dot{x}_{2_{ref}} 
		\end{align}
	for which we design the following exponentially stable \textit{desired} error dynamics
	\begin{equation}
		\dot{z}_2  + k_{P_2} z_2  = 0, \quad k_{P_2} = 50~\text{rad/s}.
	\end{equation}
	According to the results previously outlined, the incremental nonlinear dynamic inversion control law design
	follows from considering the approximate dynamics around the current reference state for 
	the dynamic equation of the pitch rate as in \eqref{eq:inc-odot}
	\begin{equation}
		\dot{q} \cong \dot{q}_{0} + \bar{g}\cdot\Delta\delta
	\end{equation}
	assuming that pitch acceleration is available for measurement and 
	the scalar $\bar{g}$ to be a factor of the accurately known estimate of $g_2$
	\begin{equation*}
		\bar{g} = k_G\cdot\hat{g}_2, \quad k_G = 1.
	\end{equation*}	

	This is rewritten in our formulation as
	\begin{equation}
		\dot{x}_2 \cong \dot{x}_{2_0} + \bar{g}\cdot\Delta u
	\end{equation}
	where recalling that $\dot{x}_{2_0}$ is an incremental instance before $\dot{x}_{2}$,
	and therefore the incremental nonlinear dynamic inversion law is hence obtained as
	\begin{equation}
		u = u_0 + \bar{g}\inv\big(\nu- \dot{x}_{2_0}\big),
	\end{equation}
	with
	\begin{equation}
		\nu = -k_{P_2} z_2 + \dot{x}_{2_{ref}},
	\end{equation}
	or more compactly
	\begin{equation}
		u = u_0 + \bar{g}\inv\big(-k_{P_2} z_2 - \dot{x}_{2_0} + \dot{x}_{2_{ref}}\big) 
	\end{equation}

	This results as desired, in the following $z_2-$dynamics
	\begin{equation}\label{z1dyn}
		\dot{z}_2 = \dot{x}_{2_0} + \bar{g}\cdot\Delta u - \dot{x}_{2_{ref}}.
	\end{equation}

	Notice that we are replacing the accurate knowledge of $f_2$ by a measurement (or an estimate)
	as $f_2\cong\dot{x}_{2_0}$, which will result in a control law which is not entirely dependent on a model, hence more robust. 
	
	We now consider these continuous-time formulations in sampled-time. To that end, we replace the small $\lambda$ with the sampling period $t_s$ so that
	$t_k = k\cdot t_s$ is the $k-$th sampling instant at time $k$, and therefore
	\begin{equation}
		\begin{split}
		u(k) = u(k-1) + \\ \bar{g}\inv\big[-k_{P_2} z_2(k-1) - \dot{x}_{2}(k-1) + \dot{x}_{2_{ref}}(k-1)\big],
		\end{split}
	\end{equation}
	where due to causality relationships we need to consider the independent variables at the same sampling time $k-1$.

	Referring back to the derived relationship between INDI and PI control, the equivalent PI control in incremental form is
	\begin{equation}
	\begin{split}
		u(k) = u(k-1) + K\cdot t_s\big[\dot{z}_{2}(k-1) + T_I^{-1}z_2(k-1)\big],
	\end{split} 
	\end{equation} 	
	with
	\begin{align}
		K = (\bar{g} \cdot t_s)^{-1}, \quad  T_I =  k_{P_2}^{-1}
	\end{align}

	The nature of the desired error dynamics (proportional) gain $ k_{P_2}$ is therefore of an integral control action, whereas the effector blending gain $\bar{g}$ act as proportional control.
	Having designed for desired error dynamics, and for a given sampling time $t_s$, tuning a pitch rate controller is only a matter of selecting a proper effector blending gain $\bar{g}$ according
	to performance requirements.
	
	{\textsl{\textbf{Remark 3}}}: Notice at this point that having the PI control in incremental form introduces a finite difference of the error state,
	which is the equivalent counterpart of what has been considered the acceleration or state derivative $\dot{x}_{2_0}$ in INDI controllers.
	
	{\textsl{\textbf{Remark 4}}}: Notice also that designing the PI control gains via INDI is highly beneficial, since only the effector blending gain is
	the tuning variable. This strongly suggests that robust adaptive control can be achieved by scheduling this variable online during flight
	and not over the whole set of gains.	
		
\begin{figure}[t]
	\centering
	\psfragfig[width=0.5\textwidth]{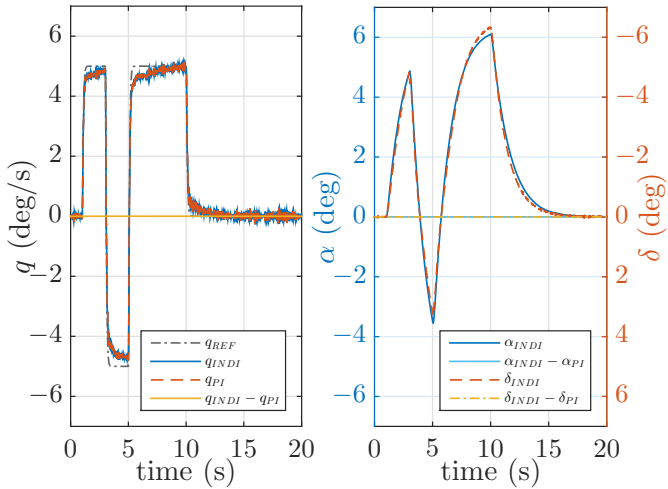}	
	\caption{INDI/PI nominal tracking control simulation 
	of the flight model \eqref{eq:example} for {$k_{P_2} = 50$ rad/s} and $k_{G} = 1$}
	\label{fig:1}       
\end{figure}

	Simulation results for the INDI/PI control are presented in Figure~\ref{fig:1},
	considering smooth rate doublets for a nominal longitudinal dynamics model at Mach 2.
	For both controllers, the same zero-mean Gaussian white-noise with standard deviation $s_{d_{q}} = 1e^{-3}$ rad/s is added to the rates to simulate noisy measurements.
	The designed INDI gains of {$k_{P_2}=50$ rad/s} and {$k_G = 1$} are 
	mapped to PI gains resulting in $K = 100~\hat{g}_2\inv$  and $T_{I} = 0.02$ s, 
	both controllers showing identical closed-loop response as expected.
		
	With this example, it is demonstrated how a self-scheduled PI can be tuned via INDI 
	by departing from desired error dynamics with the gain $k_{P_2}$, 
	and considering an accurate effector blending model estimate $\bar{g} = \hat{g}_2$.
	
\section{Conclusions}
	This paper presented a meaningful and systematic method for PI(D) tuning of robust nonlinear flight control systems 
	based on results previously reported in the robotics literature (\cite{Chang2009}) regarding the relationship
	between \textit{time-delay control} (TDC) and \textit{proportional-integral-derivative control} (PID).
	The method was demonstrated in the context of an example
	for the pitch rate tracking of a conventional longitudinal nonlinear flight model, showing
	the same tracking performance under nominal conditions.	
	
	Being incremental nonlinear dynamic inversion (INDI) equivalent to TDC 
	clearly suggests that imposing \textit{desired} error dynamics, as usual for INDI control laws,
	and then mapping these into an equivalent incremental PI(D)-controller together with control derivatives
	leads to a meaningful and systematic PI(D) gain tuning method, which is very difficult to do heuristically.

	We considered a reformulation of the plant dynamics inversion
	which reduces knowledge of the effector blending model (control derivatives) from the resulting control law, 
	reducing feedback control dependency on accurate knowledge of both the aircraft and effector blending models,
	hence resulting in robust and model-free control laws like the PI(D) control. 	
	Since usual flight control systems involves gain scheduling over the flight envelope, another key benefit 
	of this result is that scheduling only the effector blending gain seems promising for adaptive control systems.
	
\begin{ack}
M. Ruf, N. Tekles, and G. Looye are acknowledged for discussions 
leading to improvements of this paper. 
\end{ack}	

%

\bibliography{ifacconf}             

\begin{thebibliography}{16}
\providecommand{\natexlab}[1]{#1}
\providecommand{\url}[1]{\texttt{#1}}
\providecommand{\urlprefix}{URL }
\expandafter\ifx\csname urlstyle\endcsname\relax
  \providecommand{\doi}[1]{doi:\discretionary{}{}{}#1}\else
  \providecommand{\doi}{doi:\discretionary{}{}{}\begingroup
  \urlstyle{rm}\Url}\fi

\bibitem[{{Acquatella B.} et~al.(2012){Acquatella B.}, Falkena, {van~Kampen},
  and Chu}]{Acquatella2012}
{Acquatella B.}, P., Falkena, W., {van~Kampen}, E., and Chu, Q.P. (2012).
\newblock Robust {N}onlinear {S}pacecraft {A}ttitude {C}ontrol using
  {I}ncremental {N}onlinear {D}ynamic {I}nversion.
\newblock In \emph{{AIAA} {G}uidance, {N}avigation, and {C}ontrol
  {C}onference}. American Institute of Aeronautics and Astronautics, Inc.
  (AIAA-2012-4623).

\bibitem[{{Acquatella B.} et~al.(2013){Acquatella B.}, {van~Kampen}, and
  Chu}]{Acquatella2013}
{Acquatella B.}, P., {van~Kampen}, E., and Chu, Q.P. (2013).
\newblock {I}ncremental {B}ackstepping for {R}obust {N}onlinear {F}light
  {C}ontrol.
\newblock In \emph{{EuroGNC} 2013, 2nd {CEAS} {S}pecialist {C}onference on
  {G}uidance, {N}avigation, and {C}ontrol}.

\bibitem[{Bacon and Ostroff(2000)}]{Bacon1}
Bacon, B.J. and Ostroff, A.J. (2000).
\newblock Reconfigurable {F}light {C}ontrol using {N}onlinear {D}ynamic
  {I}nversion with a {S}pecial {A}ccelerometer {I}mplementation.
\newblock In \emph{{AIAA} {G}uidance, {N}avigation, and {C}ontrol {C}onference
  and {E}xhibit}. (AIAA-2000-4565).

\bibitem[{Bacon et~al.(2000)Bacon, Ostroff, and Joshi}]{Bacon2}
Bacon, B.J., Ostroff, A.J., and Joshi, S.M. (2000).
\newblock {N}onlinear {D}ynamic {I}nversion {R}econfigurable {C}ontroller
  utilizing a {F}ault-tolerant {A}ccelerometer {A}pproach.
\newblock Technical report, NASA Langley Research Center.

\bibitem[{Bacon et~al.(2001)Bacon, Ostroff, and Joshi}]{Bacon3}
Bacon, B.J., Ostroff, A.J., and Joshi, S.M. (2001).
\newblock Reconfigurable {NDI} {C}ontroller using {I}nertial {S}ensor {F}ailure
  {D}etection \& {I}solation.
\newblock \emph{{IEEE} {T}ransactions on {A}erospace and {E}lectronic
  {S}ystems}, 37, 1373--1383.

\bibitem[{Chang and Jung(2009)}]{Chang2009}
Chang, P.H. and Jung, J.H. (2009).
\newblock A {S}ystematic {M}ethod for {G}ain {S}election of {R}obust {PID}
  {C}ontrol for {N}onlinear {P}lants of {S}econd-{O}rder {C}ontroller
  {C}anonical {F}orm.
\newblock \emph{IEEE Transactions on Control Systems Technology}, 17(2),
  473--483.

\bibitem[{Chen and Zhang(2008)}]{Chen}
Chen, H.B. and Zhang, S.G. (2008).
\newblock Robust {D}ynamic {I}nversion {F}light {C}ontrol {L}aw {D}esign.
\newblock In \emph{{ISSCAA} 2008, 2nd {I}nternational {S}ymposium on {S}ystems
  and {C}ontrol in {A}erospace and {A}stronautics}.

\bibitem[{Chu(2010)}]{Chu2}
Chu, Q.P. (2010).
\newblock \emph{{A}dvanced {F}light {C}ontrol}.
\newblock Lecture notes, Delft University of Technology, Faculty of Aerospace
  Engineering.

\bibitem[{Kim et~al.(2004)Kim, Kim, and Song}]{Kim2004}
Kim, S.H., Kim, Y.S., and Song, C. (2004).
\newblock A {R}obust {A}daptive {N}onlinear {C}ontrol {A}pproach to {M}issile
  {A}utopilot {D}esign.
\newblock \emph{Control Engineering Practice}, 33(6), 1732--1742.

\bibitem[{Sieberling et~al.(2010)Sieberling, Chu, and Mulder}]{Sieberling2010}
Sieberling, S., Chu, Q.P., and Mulder, J.A. (2010).
\newblock Robust {F}light {C}ontrol {U}sing {I}ncremental {N}onlinear {D}ynamic
  {I}nversion and {A}ngular {A}cceleration {P}rediction.
\newblock \emph{Journal of Guidance, Control and Dynamics}, 33(6), 1732--1742.

\bibitem[{Simpl\'icio et~al.(2013)Simpl\'icio, Pavel, {van~Kampen}, and
  Chu}]{Simplicio2013}
Simpl\'icio, P., Pavel, M., {van~Kampen}, E., and Chu, Q.P. (2013).
\newblock An {A}cceleration {M}easurements-based {A}pproach for {H}elicopter
  {N}onlinear {F}light {C}ontrol using {I}ncremental {N}onlinear {D}ynamic
  {I}nversion.
\newblock \emph{Control Engineering Practice}, 21(8), 1065--1077.

\bibitem[{Slotine and Li(1990)}]{Slotine}
Slotine, J.J. and Li, W. (1990).
\newblock \emph{Applied Nonlinear Control}.
\newblock Prentice Hall Inc.

\bibitem[{Smeur et~al.(2016{\natexlab{a}})Smeur, Chu, and {de
  Croon}}]{Smeur2016b}
Smeur, E.J., Chu, Q.P., and {de Croon}, G.C. (2016{\natexlab{a}}).
\newblock Adaptive {I}ncremental {N}onlinear {D}ynamic {I}nversion for
  {A}ttitude {C}ontrol of {M}icro {A}ir {V}ehicles.
\newblock \emph{Journal of Guidance, Control and Dynamics}, 39(3), 450--461.

\bibitem[{Smeur et~al.(2016{\natexlab{b}})Smeur, {de Croon}, and
  Chu}]{Smeur2016a}
Smeur, E.J., {de Croon}, G.C., and Chu, Q.P. (2016{\natexlab{b}}).
\newblock Gust {D}isturbance {A}lleviation with {I}ncremental {N}onlinear
  {D}ynamic {I}nversion.
\newblock In \emph{{IEEE/RSJ} International Conference on Intelligent Robots
  and Systems (IROS)}.

\bibitem[{Smith(1998)}]{Smith1}
Smith, P.R. (1998).
\newblock A {S}implified {A}pproach to {N}onlinear {D}ynamic {I}nversion
  {B}ased {F}light {C}ontrol.
\newblock In \emph{{AIAA} {A}tmospheric {F}light {M}echanics {C}onference},
  762--770. American Institute of Aeronautics and Astronautics, Inc.
  (AIAA-98-4461).

\bibitem[{Sonneveldt(2010)}]{Sonn}
Sonneveldt, L. (2010).
\newblock \emph{Adaptive Backstepping Flight Control for Modern Fighter
  Aircraft}.
\newblock PhD thesis, Delft University of Technology, Faculty of Aerospace
  Engineering.

\end{thebibliography}
                                                   
\end{document}